\documentclass[9pt,twocolumn,twoside]{pnas-new}
\setboolean{displaywatermark}{false}
\templatetype{pnasresearcharticle} 
\usepackage{amsfonts}

\title{Directional takeoff, aerial righting and adhesion landing of semiaquatic springtails}

\author[a,b,1]{Victor M. Ortega-Jimenez}
\author[a,c,2]{Elio J. Challita}
\author[d,2]{Baekgyeom Kim}
\author[c,2]{Hungtang Ko} 
\author[d]{Minseok Gwon}
\author[d]{Je-Sung Koh}
\author[a,3]{M. Saad Bhamla}

\affil[a]{School of Chemical and Biomolecular Engineering, Georgia Institute of Technology, Atlanta, GA, USA} 
\affil[b]{School of Biology and Ecology, University of Maine, ME, USA}
\affil[c]{George W. Woodruff School of Mechanical Engineering, Georgia Institute of Technology, Atlanta, GA, USA}
\affil[d]{Dept. Mechanical Engineering, Ajou University, Republic of Korea}
\leadauthor{Ortega-Jimenez VM} 

\significancestatement{Springtails are the largest group of non-insect hexapods renowned for their unique and enigmatic appendices for jumping (furcula) and adhesion (collophore). However, it has been erroneously assumed that they are unable to control their explosive takeoff, mid-air spinning, and landing. We discover that semiaquatic springtails can indeed control all three phases of jumping by adjusting their body posture and taking advantage of their appendages. They adjust the body angle and actuation speed of their leaping organ during takeoff, change their body posture in mid-air, and exploit the hydrophilic property of their ventral adhesive tube. The combination of these strategies allows springtails to achieve locomotion control, stability, and maneuverability, which can inform the design of small bio-inspired robots with controlled landing. }

\authorcontributions{Author Contributions: VMO-J conceived and designed the study, collected animals, performed animal experiments; VMO-J, HK, EJC, and MSB analyzed data; HK conceived theoretical model on directional jumping.  EJC conceived theoretical model and simulation on landing impact on the water; BK,MG, and J-SK designed jumping robot and analyzed data;  VMO-J, HK, EJC, EC, MSB, BK, and J-SK interpreted the results and wrote the paper.}
\authordeclaration{The authors declare no competing interest.}
\equalauthors{\textsuperscript{2}E.J.C., B.K. and H.K.  contributed equally to this work.}
\correspondingauthor{\textsuperscript{1,3}To whom correspondence should be addressed. E-mail: vortex@maine.edu; saadb@chbe.gatech.edu}


\begin{abstract}
Springtails (Collembola) have been traditionally portrayed as explosive jumpers with incipient directional takeoff and uncontrolled landing. However, for these collembolans who live near the water, such skills are crucial for evading a host of voracious aquatic and terrestrial predators. We discover that semiaquatic springtails \textit{Isotomurus retardatus} can perform directional jumps, rapid aerial righting, and near-perfect landing on the water surface. They achieve these locomotive controls by adjusting their body attitude and impulse during takeoff, deforming their body in mid-air, and exploiting the hydrophilicity of their ventral tube, known as collophore. Experiments and mathematical modeling indicate that directional-impulse control during takeoff is driven by the collophore’s adhesion force, the body angle, and the stroke duration produced by their jumping organ, the furcula. In mid-air, springtails curve their bodies to form a U-shape pose, which leverages aerodynamic forces to right themselves in less than $\sim$20 ms, the fastest ever measured in animals. A stable equilibrium is facilitated by the water adhered to the collophore. Aerial righting was confirmed by placing springtails in a vertical wind tunnel and through physical models. Due to these aerial responses, springtails land on their ventral side $\sim$85\% of the time while anchoring via the collophore on the water surface to avoid bouncing. We validated the springtail biophysical principles in a bioinspired jumping robot that reduces in-flight rotation and lands upright $\sim$75\% of the time. Thus, contrary to common belief, these wingless hexapods can jump, skydive and land with outstanding control that can be fundamental for survival.
\end{abstract}

\dates{This manuscript was compiled on \today}
\doi{\url{www.pnas.org/cgi/doi/10.1073/pnas.XXXXXXXXXX}}

\begin{document}

\maketitle
\thispagestyle{firststyle}
\ifthenelse{\boolean{shortarticle}}{\ifthenelse{\boolean{singlecolumn}}{\abscontentformatted}{\abscontent}}{}

\dropcap{S}pringtails (Arthropoda:Collembola) are the most widespread, abundant and diverse group of non-insect hexapods on the planet, which are known for their major role in soil ecology, and unique adaptions to catapult themselves into the air \cite{hopkin1997biology}. Collembolans' jumping performance has been extensively studied in terms of locomotion \cite{christian1978jump,bush2006walking,sudo2015jumps}, morphology \cite{sudo2013observations,sudo2015jumps,chen2019structure}, behavior \cite{bauer1987habitat, haagvar2010review,zettel2013jumping}, energetics \cite{ruhfus1995investigations}, and computational modeling \cite{brackenbury1993jumping}. It has inspired the design of mechanical jumpers \cite{hu2010water,sudo2013kinematics} and robots \cite{ma2021biologically}. Previous biomechanical studies suggested that springtails’ jumping, and particularly their landing, are uncontrollable and unpredictable \cite{christian1978jump,hu2010water}, given that these wingless arthropods can reach impressive body rotations in mid-air ($\sim$500 Hz, see \cite{sudo2013kinematics}). In contrast, behavioral and ecological studies indicate that these tiny arthropods can perform sophisticated maneuvers, navigation and consistent landing. For example, directed leaping and controlled landing have been observed during the massive and long-distance migration (up to 300,000 bodies/day) of snow-dwelling springtails \cite{hagvar1995long}. Furthermore, scanning microscope images from a recent report suggest that the collophore may be used for adhesion to the water surface, cleaning and nutrient absorption \cite{chen2019structure}. It can play a key role in controlling takeoff direction and trajectory \cite{favret2015adhesive}. Similarly, a previous mathematical analysis and computational model suggests that varying the furcula's length  may influence the vertical and horizontal range reached during jumping \cite{brackenbury1993jumping}. Despite these observations and speculations, it is unclear how springtails are able to control their jumping and landing by using the collophore, the furcula and their slim bodies (i.e., entomobryomorpha).

Aerial righting is a broadly used strategy for in-flight control. It is exploited by animals, as well as by wind-dispersed seeds \cite{cummins2018separated,ortega2019superb} to gain a favorable orientation in mid-air and consequently during landing \cite{jusufi2011aerial}. Self-righting has been studied in wingless mammals \cite{McDonald}, reptiles \cite{jusufi2008active}, insects \cite{ribak2013adaptive, zeng2017biomechanics, yanoviak2009gliding, yanoviak2010aerial}, and arachnids \cite{yanoviak2015arachnid}, but it has never been reported in non-insect hexapods such as springtails. In general, those studies indicate that self-righting is size dependent. Large animals, such as cats or geckos, recover from an upside-down posture by using inertial responses of their bodies, limbs or tail \cite{jusufi2008active, jusufi2011aerial}. In contrast, insects, such as larval stick insects and adult locust, rely on aerodynamic responses produced by their appendages to correct their body orientation \cite{zeng2017biomechanics,FaisalMatheshon2001}. Accordingly, it stands to reason that millimeter-sized animals, such as springtails, may exploit aerodynamic forces to recover from an unfavorable upside-down orientation. Nevertheless, aerial righting in springtails can be dynamically more challenging than free fall righting, because collembolans are required to suddenly reduce their fast body spinning in mid-air in order to gain a favorable orientation during the collision with the surface. Shock absorption and attachment with the water surface during landing is also crucial to performing a subsequent controlled jump, which seems challenging even for winged insects \cite{Reicheletal2019}. Without such mechanisms, springtails would bounce uncontrollably, causing potential physical damage, as well as increasing the likelihood to be targeted and captured by predators.

In this contribution, we investigate how wingless springtails maintain a tight control for takeoff, in mid-air and during landing. We focus on semiaquatic springtails \textit{ Isotomurus retardatus} \cite{folsom1937nearctic} that live on the water surface, for which locomotion control and maneuvering are essential for survival, given the persistent pressure that springtails experience against countless predators \cite{hopkin1997biology}. We use high-speed videography, kinematic analysis, mathematical modeling, particle image velocimetry, a vertical wind tunnel, and biomimetic robophysical models to investigate the locomotion control abilities of these millimeter-sized wingless arthropods that live on the air-water interface. In addition, we designed a bio-inspired robot with the capacity to control its landing for enabling stable repetitive jumps based on our findings of springtails' aerial righting.

\begin{figure*}
	\centering
	\includegraphics[width=1\linewidth]{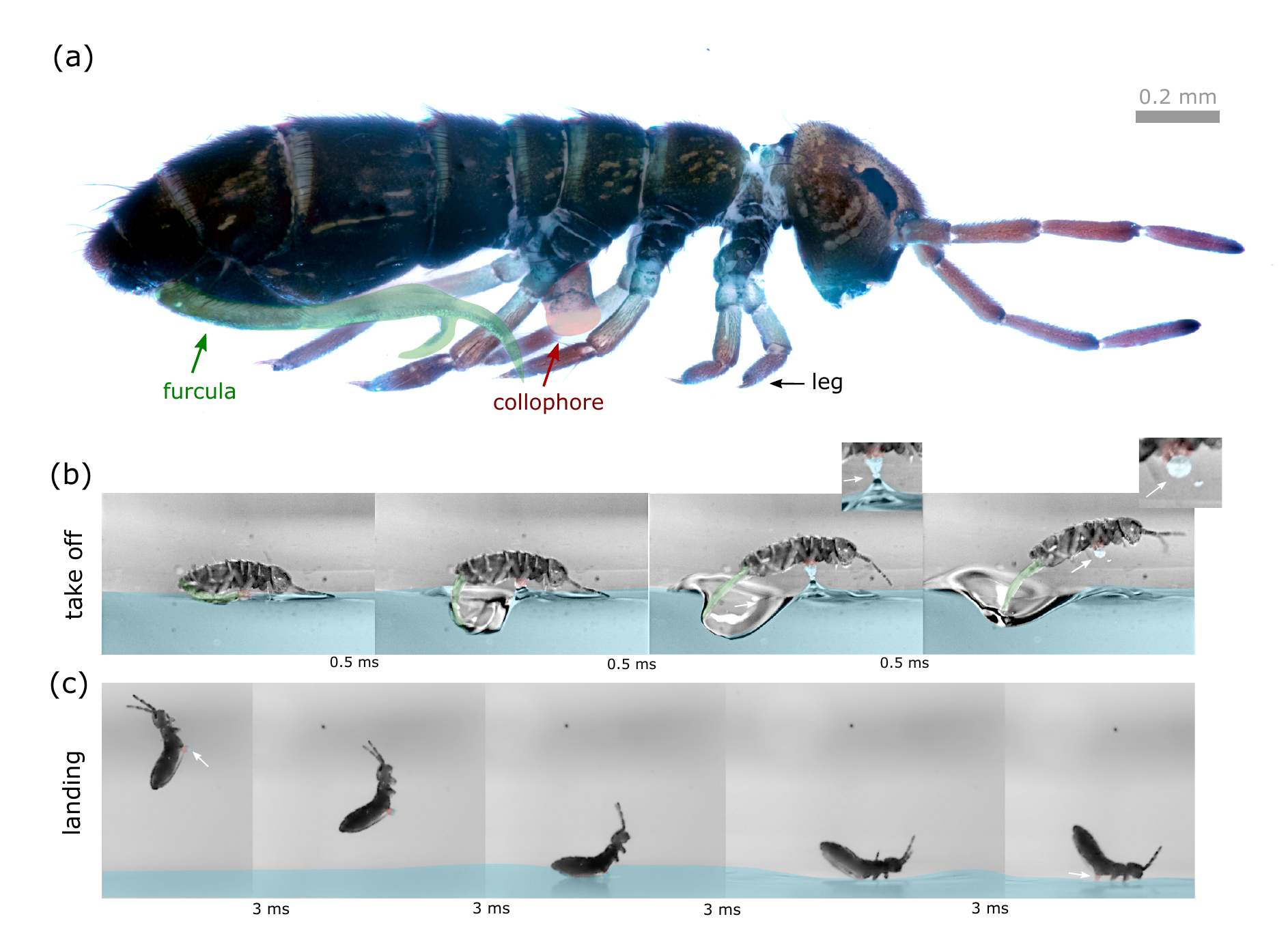}
	\caption{\textbf{Semiaquatic springtail\textit{ Isotomorus retardatus}.} (a) Notice the manubrium-furcula and collophore highlighted in green and red, respectively. Image composites are from recordings of springtails taking-off (b) and landing (c). \textbf{(b)} Detail of water adhesion of the collophore and the droplet collected after detaching from the water surface are shown above their respective frame. \textbf{(c)} Notice that during a successful landing, springtails attach to the water surface using the collophore.} 
	\label{figure_1}
\end{figure*}

\section*{Results}

%

\begin{figure*}
 	\centering
	\includegraphics[width=1\linewidth]{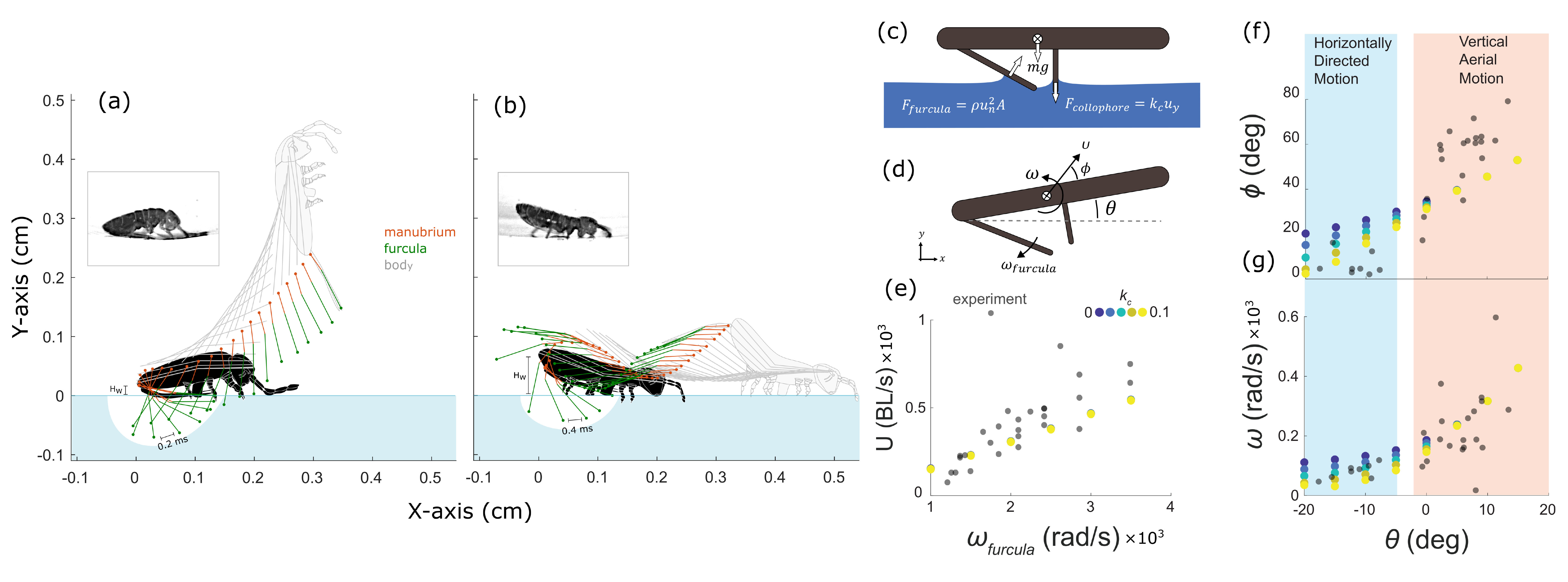}
	\caption{\textbf{Trajectories and theoretical model of springtails’ catapulting from the water surface.} \textbf{(a)} Springtail presenting a fast vertical takeoff. Body, manubrium and furcula are highlighted in grey, orange and green. Notice that the individual humped their body before jumping and thus the tip of the abdomen is almost touching the water. \textbf{(b)} Springtail presenting a slow and horizontal takeoff. Notice that the tip of the abdomen is maintained farther from the water surface. \textbf{(c)} springtail takeoff model is affected by furcula propulsion, collophore adhesion, and gravity. \textbf{(d)} definitions of the metrics used in takeoff experiments and simulations. \textbf{(e)} takeoff speed $U$ as a function of furcula opening velocity $\omega_{furcula}$. \textbf{(f)} takeoff angle $\phi$ and (g) angular velocity $\omega$ of the springtail jumping from various initial orientation $\theta$. Color dots are simulation data from the model, with color representing the intensity of collophore cohesion, controlled by coefficient $k_c$. Gray dots are experimental measurements ($N=28$).See text for details. } 
	\label{figure3_takeoff_model}
\end{figure*}

\subsection*{Takeoff control}
\begin{figure*}
	\centering
	\includegraphics[width=1\linewidth]{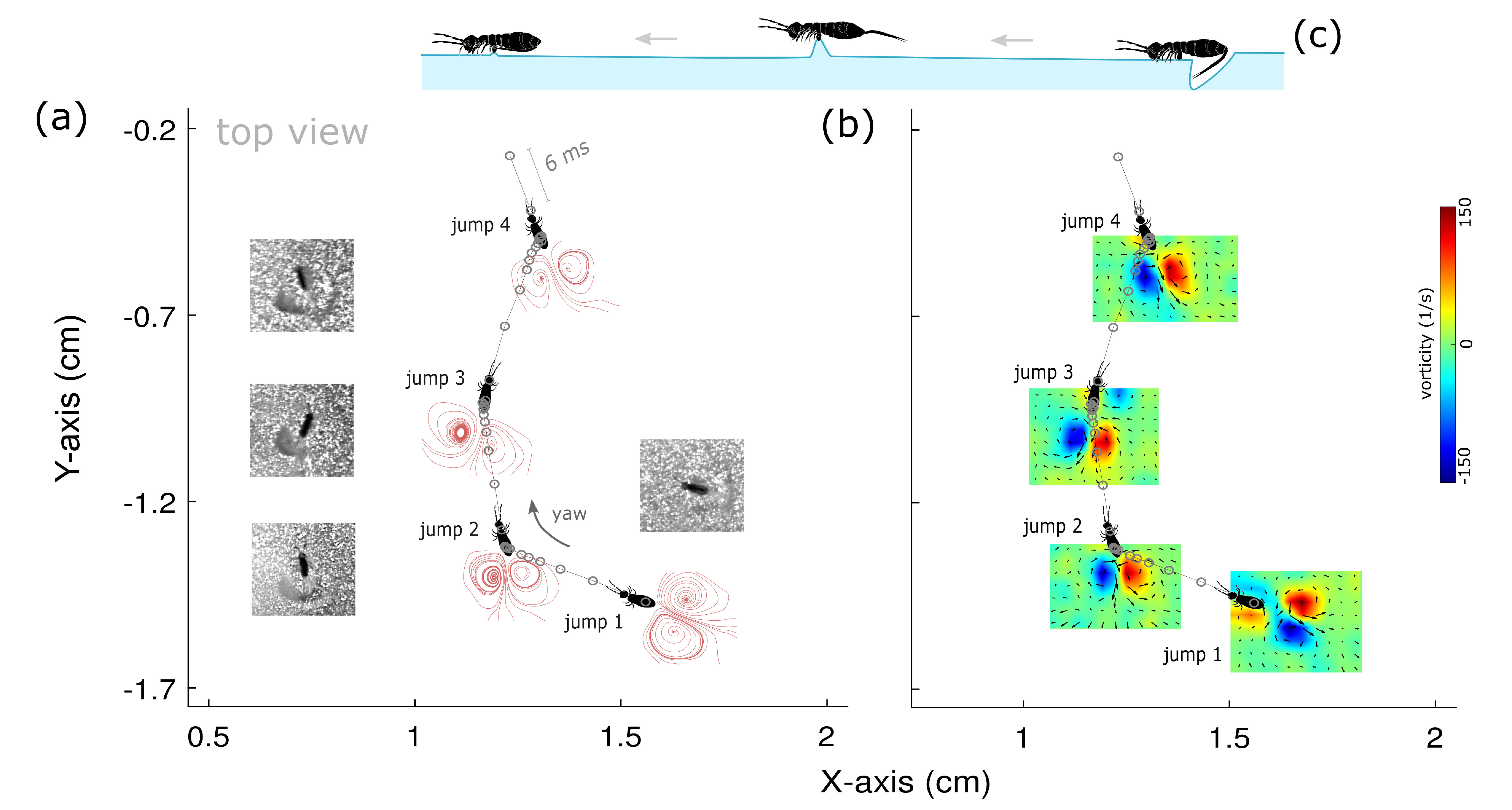}
	\caption{\textbf{Springtail in Horizontally Directed Motion (HDM).} \textbf{(a)} Streamlines and \textbf{(b)} vorticity fields from particle image velocimetry analyses. Average traveling speed was 28 cm/s ($\sim$280 bodies/s), which is similar to that of juvenile water striders during rowing \cite{ortega2017escape}. Time interval between grey circles is 6 ms. Notice in picture frames the wave produced by springtails during jumping. Yaw turning (grey curved arrow in (a)) is achieved by asymmetric leg movement on the water. \textbf{(c)} Drawing shows a hypothesised mechanism of translation motion during each horizontal jump.} 
	\label{figure_3}
\end{figure*}

Before jumping, springtails anchor on the water surface using the collophore (Figure \ref{figure3_takeoff_model}, Video 1). We found that some individuals humped their body before jumping, which lowers the tip of the abdomen, and consequently, the furcula (Figure \ref{figure3_takeoff_model}a). During takeoff, the hydrophilic collophore traps a small water droplet visible in the high-speed images as shown in Figure \ref{figure_1}b, that plays an important function in both aerial righting and landing as described in later sections. After takeoff, these humped individuals jumped with vertical trajectories and high speeds. In contrast, other individuals kept their body straight and elevated the tip of the abdomen from the water (Figure \ref{figure3_takeoff_model}b). In this case, springtails’ trajectories were horizontal and much slower.
In a few extreme cases, we even observed individuals locomote on the water surface without detaching the collophore from the surface. These springtails were able to skip on the water surface, frequently actuating their furcula. They reached traveling speeds of up to 28 cm/s ($\sim$280 bodies/s), as well as produced a vortical wake with a flow velocity and vorticity of up to 50 cm/s and 150 1/s, respectively (Figure \ref{figure_3}ab, S1, Video 1). We observed that springtails moving horizontally on the water use their legs to adjust their yaw direction before each leaping (Video 1).

To quantify these observations, we digitized 27 high-speed recordings of springtail jumps. We found a linear relationship between the body angle at the starting of takeoff $\theta$ and the height of the abdomen normalized by body length $H_{abdomen}$ ($\theta$=17-7$\times H_{abdomen}$, $r^2$ =0.82, $F_{1,25}$=113, $p \ll 0.001$), showing that the insect can change $H_{abdomen}$ through changing their angle $\theta$. $\theta$ also dictates the takeoff kinematics (Figure S2). The takeoff angle $\phi $ ( $\phi $ =39+2.6$\theta$, $r^2$ = 0.83, $F_{1,25}$ = 124, $p \ll 0.001$), and the takeoff speed $u$ ($u$=381-16$\times \theta$, $r^2$ = 0.57, $F_{1,25}$ = 33.5, $p \ll 0.001$), were linearly related with the body angle $\theta$.  Furthermore, takeoff speed $u$ depends on the furcula’s angular speed ($u$=-161+0.3$ \times \omega_{furcula}$, $r^2$ =0.74, $F_{1,25}$=62, $p \ll 0.001$).  In contrast, the average angular rotational speed of the body $\omega$ was weakly related with $\theta$ ($\omega$ =180+9$\times \theta$, $r^2$ =0.4, $F_{1,25}$=17, $p \ll 0.001$)(Figure S2). Reynolds number based on body thickness corresponds to $\sim$14.

Can springtails control their jumping pose to achieve their desired directions and speeds? To understand the takeoff mechanism, we constructed a mathematical model that predicts the motion of the insect based on Newtonian mechanics. The external forces that drives the action includes propulsion from the furcula, adhesion from the collophore, and gravity (see details in the method section). We study the effects of body orientation $\theta$, furcula opening velocity $\omega_{furcula}$, and collophore adhesion $k_c$.

Figure \ref{figure3_takeoff_model}d demonstrates that when the body pose $\theta$ is fixed, the model predicts that takeoff velocity $U$ increases with the furcula velocity $\omega_{furcula}$, fitting the experimental measurement qualitatively. Figure \ref{figure3_takeoff_model}f-g shows that, when furcula velocity $\omega_{furcula}$ is fixed, jumpers can indeed control their takeoff kinematics through controlling body pose $\theta$. 

Despite the relatively smooth transition, we can identify two regimes: Horizontally Directed Motion (HDM) can be achieved with negative $\theta$; Vertical Aerial Motion (VAM) at large positive $\theta$. Springtails that adopt HDM, such as the one shown in Figure \ref{figure3_takeoff_model}b, take off at an angle closer to the water surface and with less rotation. We surmised that the collophore adhesion originates from the thin-film drainage process around the collophore. Therefore, the adhesion is proportional to the collophore retraction velocity and the water film thickness. We lumped the impact of film thickness and water viscosity to a coefficient $k_c$. While this coefficient can not be controlled by the springtail, studying the influence of this parameter can inform the role of collophore in the jumping mechanics. For HDM, the presence of collophore allows the insect to take off even closer to the surface and rotate even less. In fact, the collophore adhesion is required for springtails to "skip" on the water surface as shown in Figure \ref{figure3_takeoff_model}. On the other hand, springtails that adopt VAM jump at higher transitional and angular speeds, as well as directed vertically (Figure \ref{figure3_takeoff_model}a). Both regimes were found in experiments and validated through our simple mathematical model. Together, we show that springtails can control their takeoff speeds through controlling their furcula opening speeds, and they can control their jumping direction and spinning through controlling their orientation. Figure \ref{figure3_takeoff_model}e-g shows that the collophore doesn't affect the takeoff speed, direction or spinning for VAM (large $\theta$). We observed that only one individual exhibited a very small body rotation while jumping vertically (VAM)(Figure \ref{figure3_takeoff_model}g, Video S1).


\subsection*{Collophore’s adhesion and droplet capture}

The collophore’s hydrophilic adhesion contributes to the HDM, and traps a small droplet that is fundamental for self-righting post-takeoff. To quantify this adhesion force and mass of droplets captured, we perform an experiment by placing individual springtails on a rotating disk (n=23)(Figure S3). Data is presented as the average value $\pm$ one SD. Individuals were ejected at an acceleration of 50$\pm$30 m/s$^2$. Ejection speed was 1.1$\pm$0.4 m/s (Figure S3c). Using the acceleration during ejection and the body mass, we estimated that the collophore attaches to a surface with an adhesion force of $~$7$\pm$4 $\mu N$. The amount of water collected by the collophore using the dimensions of collophore and the attached droplet corresponded to $\sim$3\% of the body mass (Figure \ref{figure_1}b, S3).

\begin{figure*}[!ht]
\centering
\includegraphics[width=0.7\linewidth]{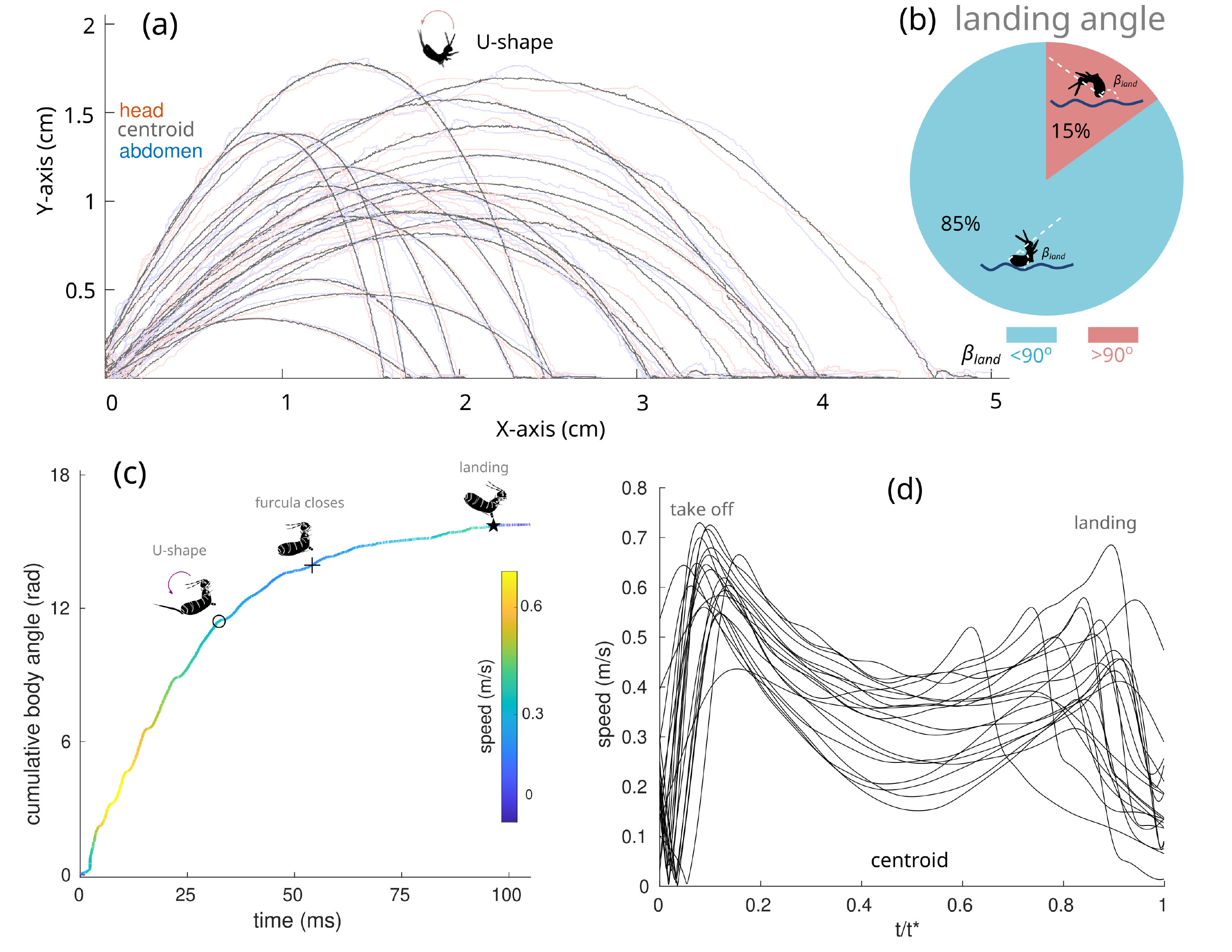}
\caption{\textbf{Vertical Aerial Motion (VAM) and near-perfect landing of springtails.} \textbf{(a)} Full trajectories of 20 springtails during jumping and landing. Head, centroid and abdomen tip are highlighted in orange, grey, and blue, respectively. \textbf{(b)} Springtail's landing angle with respect to the water surface. Lower than 90 degrees indicates a ventral landing; larger than 90 degrees, a dorsal landing. \textbf{(c)} Cumulative body angle over time of a springtail effectively reducing and stopping body rotations by deforming its body to a U-shaped posture. Note that the furcula closing seems to have no effect on body rotation. \textbf{(d)} Centroid speed ($\sqrt{\upsilon_x ^2 + \upsilon _y^2}$) over normalized time (i.e., total jumping duration is one) of the 20 springtails shown in a. Two Peaks on curves represent takeoff and landing, respectively.} 
\label{figure_5}
\end{figure*}

\subsection*{Flight trajectories and landing success}
We zoom out to discuss springtails' flight trajectories post takeoff. Springtails (N=20) during jumping reached a horizontal and vertical distance of 3$\pm$1 cm (36$\pm$12 BL) and 1.1$\pm$0.3 cm ($\sim$12$\pm$4 BL), respectively (Figure \ref{figure_5}a). Takeoff angle, maximal body rotational frequency and flight duration were 47$\pm$10$^\circ$, 71$\pm$42 Hz, and 93$\pm$17 ms, respectively. Speed of the center of gravity during landing (48$\pm$9 cm/s) was reduced 25\% in comparison with that during takeoff (63$\pm$7 cm/s). After launching, individuals deformed the bodies into a U-shape at 15$\pm$6 ms and the furcula closed in mid-air at 26$\pm$11 ms (Figure \ref{figure_5}c). We found that 85\% of the sampled springtails landed on their ventral side, despite the high body spinning frequency in mid-air after takeoff (Figure \ref{figure_5}b). Cumulative angle time series indicate that springtails adopted an early U-shape in mid-air reduced their rotational speeds (Figure \ref{figure_5}c) and corrected body orientations to prepare for ventral landing.    

\subsection*{Aerial righting}

\begin{figure*}[!ht]
	\centering
	\includegraphics[width=.8\linewidth]{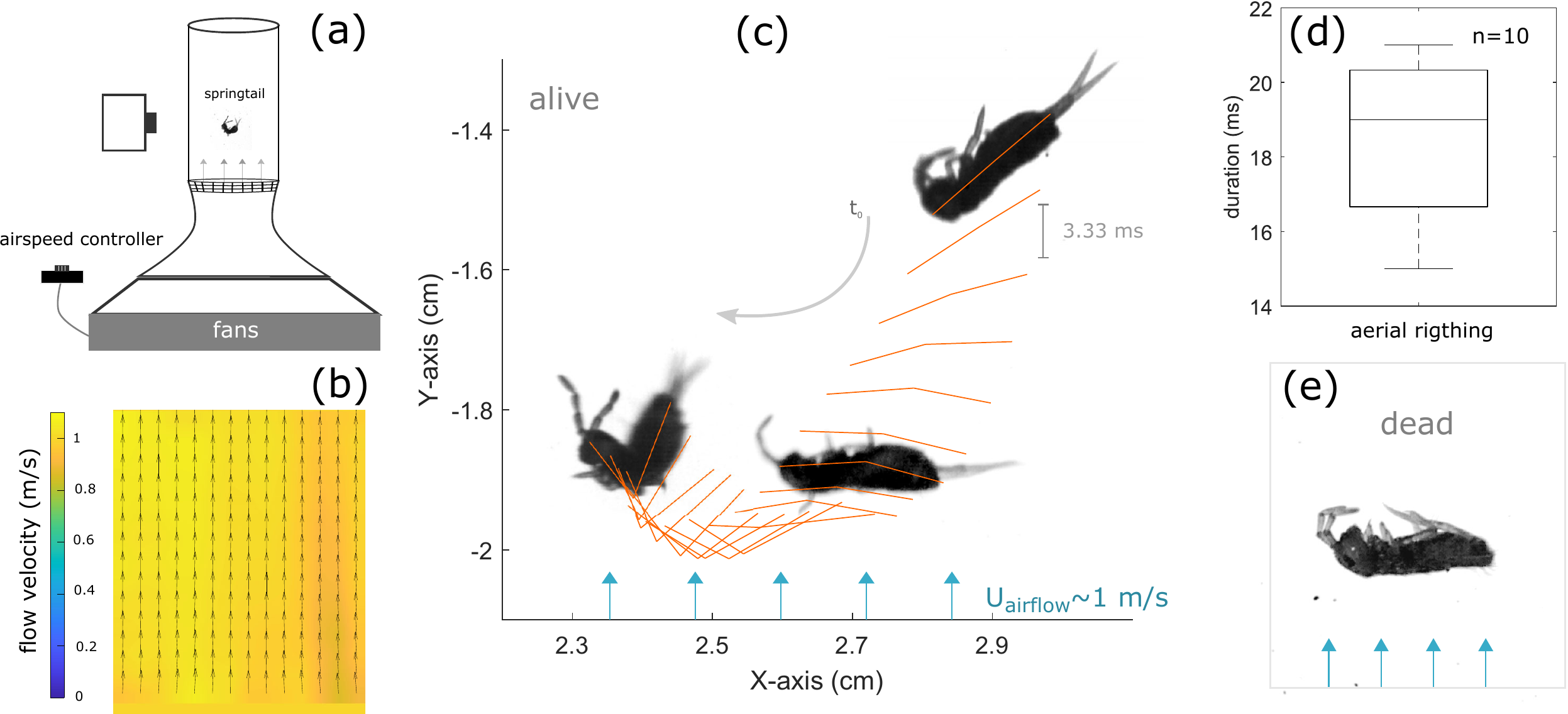}
	\caption{\textbf{Springtail aerial righting in wind tunnels.} \textbf{(a)} Diagram showing experiments of springtails aerial responses in a vertical wind tunnel (not to scale). \textbf{(b) }Velocity fields of the airflow inside the wind tunnel. \textbf{(c)} Alive springtail recovering from an upside-down posture after curving its body. \textbf{(d)} Aerial righting duration (n=10). Box plot showing median (dark line), 25th–75th percentile (box) and extreme values (whiskers). \textbf{(e)} Dead springtails keep an upside-down posture inside the vertical wind tunnel.}
	\label{figure_6}
\end{figure*}

To test this self-righting hypothesis, we placed both live and dead springtails in a vertical wind tunnel (flow speed$\sim$1 m/s) (Figure \ref{figure_6}, Video S3). In both conditions, the animal started their fall with their back facing vertically downwards. Live individuals flipped their orientation with their ventral side pointing down immediately after adopting a U-shape. This aerial righting happened in less than $\sim$20$\pm$2 ms (N=10), the fastest ever measured in animals as far as we know. Furcula remained extended during the aerial maneuvers while the antennae retracted backwards, and the legs were extended. We also corroborated a reduction and cease of body rotation after springtails curved their bodies in the wind tunnel experiments, which agrees with the reduction of accumulative body angle over time observed in jumping trials (Figure \ref{figure_5}c, Video S3).

\subsection*{Free fall physical models}
\begin{figure*}[!ht]
	\centering
	\includegraphics[width=.8\linewidth]{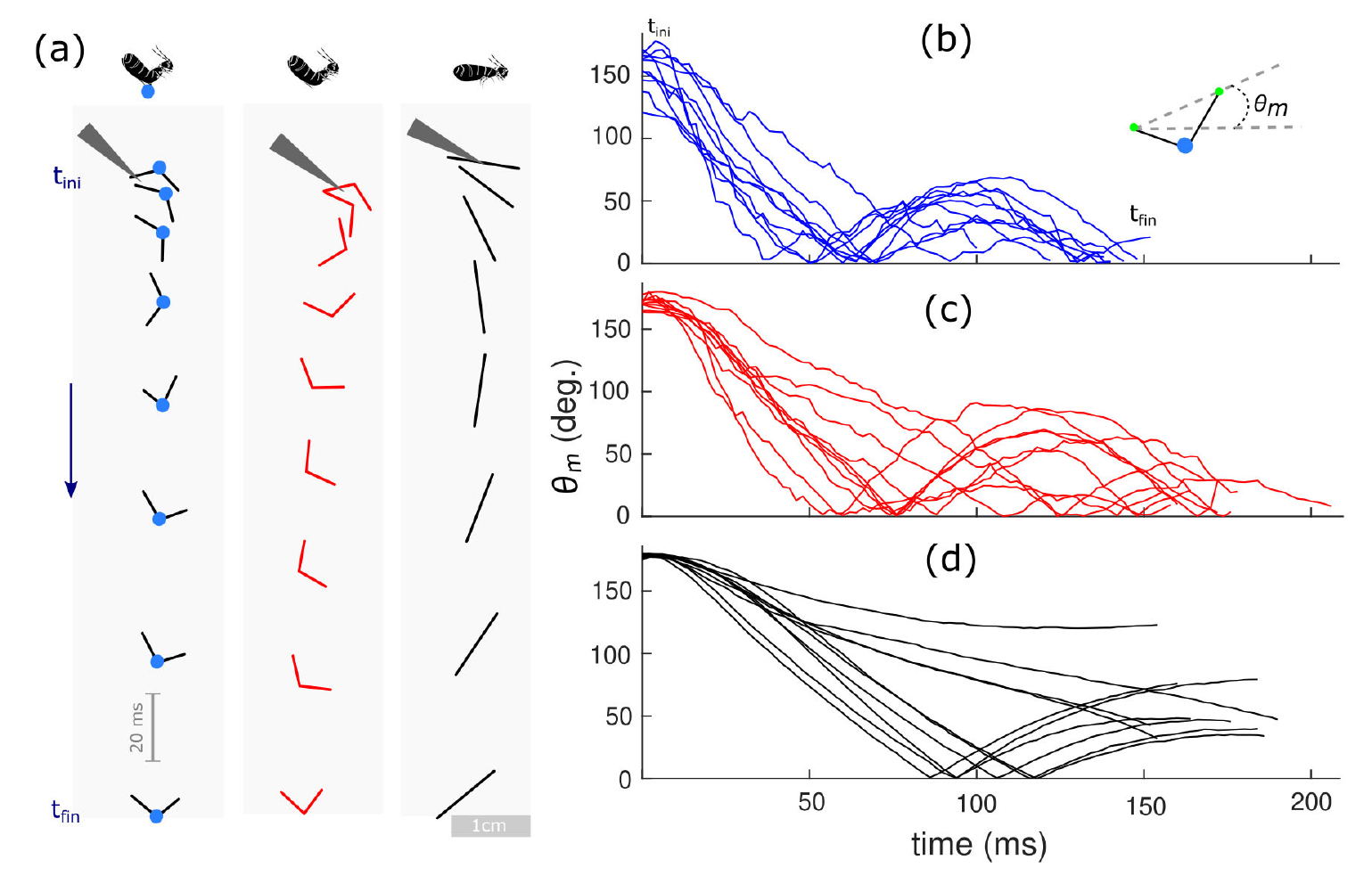}
	\caption{\textbf{Physical models during free fall (n= 10 for each case)}. \textbf{(a) } Free fall trajectories from a video of a U-shaped strip with a droplet (left), U-shaped strip (center) and flat plate (right). \textbf{(b) }Angle over time of a U-shaped strip with a droplet on its vertex. \textbf{(c)} Angle over time of a U-shaped strip without a droplet. \textbf{(d)} Angle over time of a flat strip. See text and Video S3 for details.}
	\label{figure_7}
\end{figure*}

In order to understand how the U-shape and water droplet influence springtails' righting we performed free fall experiments using 3 plastic film models (U-shape with droplet, U-shape without droplet, and flat strip)(Figure \ref{figure_7}a). Both U-shape plastic strips corrected passively their upside-down position in mid-air to finally land on its vertex (n=10 each) (Figure \ref{figure_7}). Nevertheless, faster righting and smaller angle variation during rotation was observed in U-shaped strips with the droplet than without it. In contrast, flat strips in free fall landed at different angles on the ground. These results support that body deformation and the droplet collected by the collophore help the animal recover from an upside-down posture. Reynolds number based on strip thickness corresponded to 22.

\subsection*{Springtail-inspired self-righting robot}
\begin{figure*}[!ht]
	\centering
	\includegraphics[width=.8\linewidth]{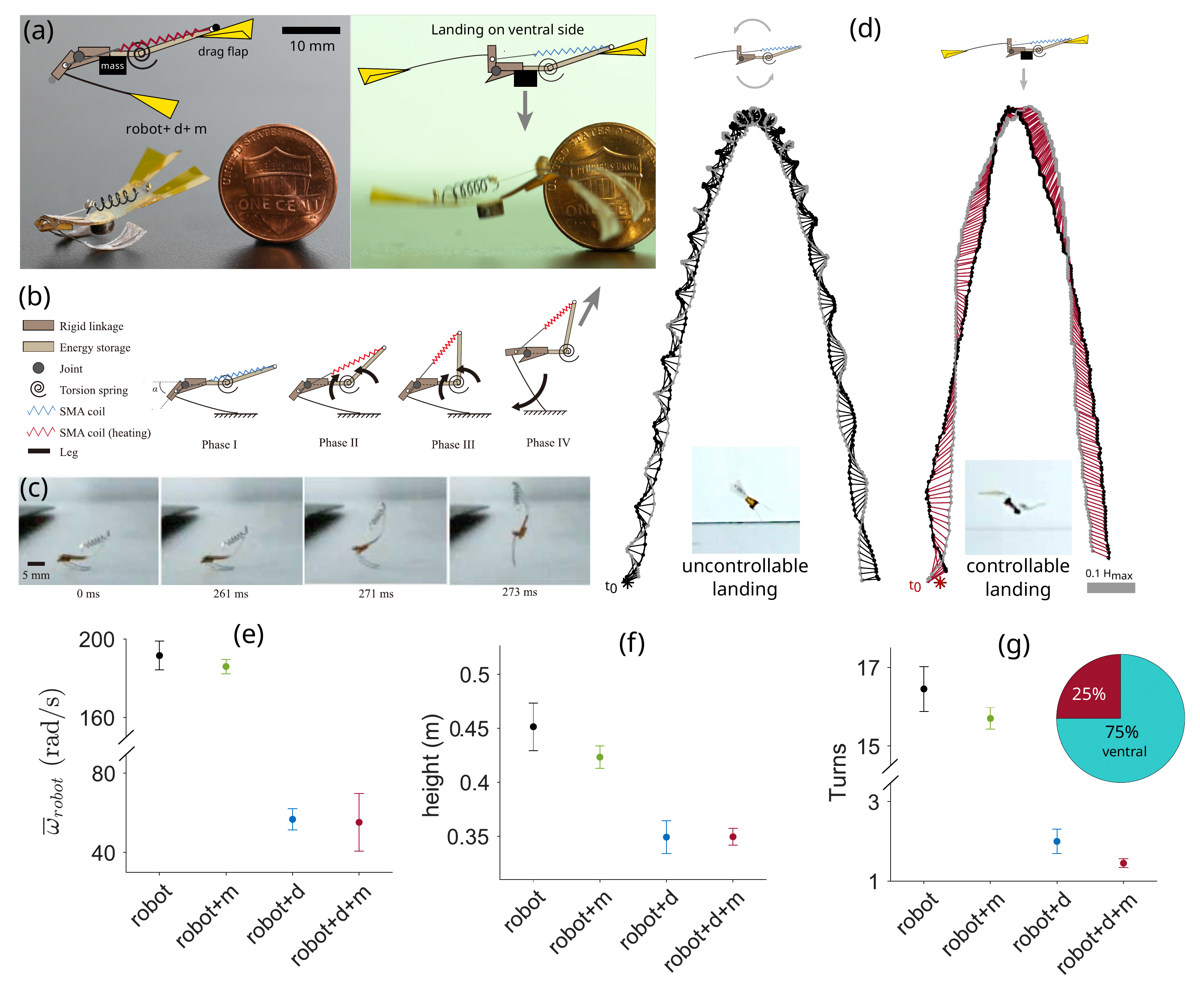}
	\caption{\textbf{Springtail-inspired self-righting robot.} \textbf{(a)} Robot with drag flaps + extra mass during take off and landing. \textbf{(b)}  The schematic diagram of the jumping mechanism from Phase I to Phase IV. \textbf{(c)} Sequential picture of the jumping motion. \textbf{(d)} Normalized trajectories with maximal height of a robot without (black) and with drag flaps+extra mass (red). Horizontal lines represent body orientation. \textbf{(e)} Pie chart represents frequency of successful and failed ventral landing (N=20). Angular speed \textbf{(e)}, maximal height \textbf{(f)} and number of turns \textbf{(g)} of robot jumping in four treatments (N=5 for each trial). Error bars represent average value $\pm$ one standard deviation.}
	\label{figure8}
\end{figure*}

We designed a jumping robot (86 mg) that mimics the catapulting mechanism of springtails, which consisted of a lightweight structure and actuator to realize jumping functionality (Figure~\ref{figure8}c). The robot is activated by heating a shape memory alloy (SMA) coil which produces tension and a sudden opening of the leg (furcula-like structure), catapulting the robot into the air (Figure~\ref{figure8}ab). Before any springtail-inspired design strategy is implemented, the baseline robot rotates uncontrollably during its flight trajectory. Can one improve the robot's performance based on our discoveries in springtails' righting principles (see Video S2)?

We made three modifications to the robot to explore how drag (U-shape) and extra mass (droplet) affect landing performance. Five jumping trials of each modification (treatments), as well as only the robot, were filmed at 1000 frames/s. Modifications were as follows: robot with an extra mass (98 mg)($robot+m$), robot with drag flaps (98 mg) (robot$+$d), and robot with added mass plus drag flaps (110 mg)($robot+m+d$)(Figure~\ref{figure8}c) (Video S2). Kruskal–Wallis tests were significant between at least two treatments for rotational speed ($\chi^2_3=15, P<0.01$), maximal height ($\chi^2_3=15.6, P<0.01$), and number of turns ($\chi^2_3=17.3, P<0.001$). Robot with drag enhancers and mass (robot+m+d) during jumping showed significantly lower average rotational speed, maximal height (Figure S4) and smallest total number of turns (4, 0.35 and 11 times, respectively) in comparison with the robot without additions ($P<0.01$ for all pair contrasts)(Figure~\ref{figure8}d-g). We note that the robot without any modifications ($robot$), as well as with the extra mass ($robot+m$) rotated uncontrollably during its flight trajectory, even during its descent and landing. In contrast, both the robot with drag enhancers ($robot+d$) and with drag enhancers plus extra mass ($robot+m+d$) descended smoothly with its ventral side pointing downwards. Additionally, we found that the robot with drag flaps and extra mass ($robot+m+d$) landed on its ventral part $75\%$ of the time ($n=20$). Thus, aerodynamics torque and an extra mass are effective to reduce rotation and facilitate a controllable landing in small robots, which enables them to jump repeatedly without an extra righting strategy.

\subsection*{Adhesive landing on the water surface}
We have discussed so far how the springtails harness their morphology, especially the collophore, to influence their take-off from the air-water interface and orientation mid-air. Next we demonstrate that this hydrophilic structure also plays a role in damping their landing, back on the surface of water. We observe that individuals approach the water surface with their ventral part directed downwards, which subsequently attaches to the water surface with the collophore. This action effectively absorbs the momentum during the impact (Figure \ref{figure_9}) by producing capillary waves (Video S2). In contrast, individuals landing on their backs or laterally, bounced uncontrollably on the water, until they manage to correct their position on the water surface using their legs. To confirm the role played by the capillary adhesion of the collophore (through droplet), we removed the water from the collophore from a few individuals and allowed them to land on a dry solid substrate. Despite landing ventrally on the dry collophore, they bounced several times due to inability of the collophore to stabilize their landing (Figure S5). Individuals landing on their backs on the water surface took $\sim$10 times longer to correct their position and adhere at the interface with the collophore, than those that landed ventrally in the first place (44 ms vs. $\sim$4 ms, respectively).

\subsection*{Landing model}
\begin{figure*}[!ht]
	\centering
	\includegraphics[width=.8\linewidth]{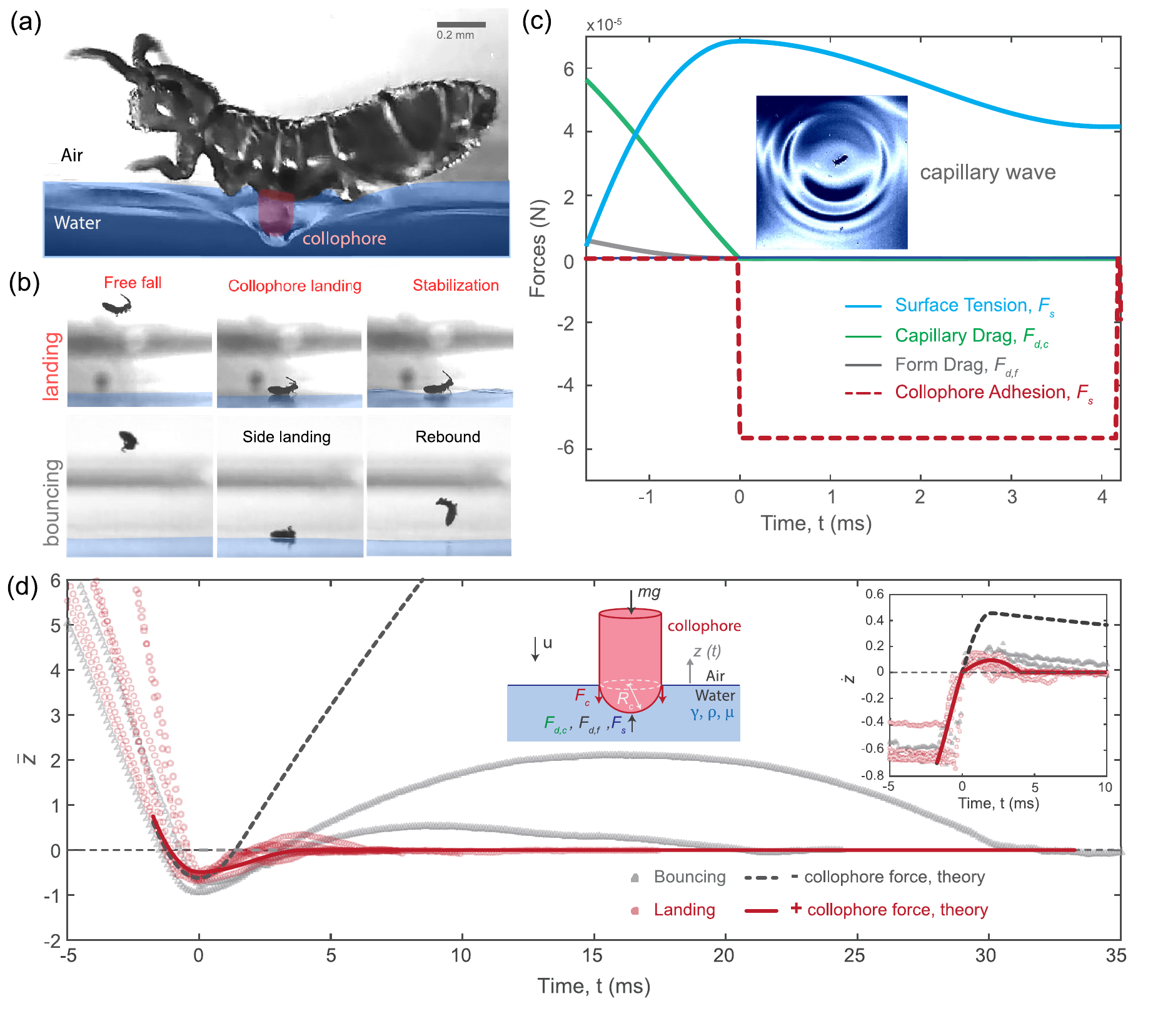} 
	\caption{\textbf{Springtails adhesive landing on the air-water interface}
	\textbf{(a)} Before impact, springtails arch their body further extending their collophore as they make contact with the water surface. During impact,
    the collophore is the main structure that interacts with the interface. \textbf{(b)} The orientation of springtail's body is key to the success of its landing. Landing is controlled when the collophore is positioned to adhere to the water surface as quickly as possible. Capillary waves are observed post-impact as the organism oscillates to equilibrium. Alternatively, if the springtail lands in an other configurations (side or back), we observe that the springtail rebounds uncontrollably.\textbf{(c)} The evolution of hydrodynamic forces over time. Note that collophore adhesion acts to slow the rebounce after the springtail reaches maximum depth and starts moving upwards. \textbf{(d)} We develop a simplified mathematical model to study the effect of collophore adhesion on the landing dynamics of springtails. Hydrodynamic forces generated consist of weight, drag, surface tension and collophore adhesion. Experimental and theoretical results of the normalized displacement $\bar{z}=(z(t)-z_{t\rightarrow \infty})/D_s$ show that collophore adhesion is important to quickly arresting the landing dynamics of the springtail. Without collophore force, the springtail would fly off the surface of the water.} 
	\label{figure_9}
\end{figure*}

How do springtails land on the water surface and minimize any unwanted bouncing and uncontrolled tumbling? We analyze the landing dynamics of springtails and compare them to a reduced-order hydrodynamic model. Springtails land at the water surface with an impact speed of $u_i = 0.54 \pm 0.12\;m/s\;(N=9)$, reaching a maximum depth of $z_{max}\sim 0.2\;mm$ (or $\sim 60 \%$ the body width $D_s$) under the water surface for a duration of $\sim 3-5\;ms$. As discussed previously, the outcome of landing depends highly on the morphology of the springtails right before impact. When springtails land on their ventral side (denoted by `collophore landing'), their exposed collophore due to their arched U-shaped morphology is the first to interact with the water surface (Figure \ref{figure_9}a). In this case, springtails quickly stabilize on the water surface and remain anchored (Figure \ref{figure_9}b). In contrast, when springtails land on their peripheral or dorsal side, they bounce off the surface of the water.  
 We model the dynamics of the collophore at the water-air interface mathematically (See Materials and Methods). The collophore geometry is approximated as a cylinder with a hemispherical end having a radius $R_c$ (Diameter $D_c$)(Figure \ref{figure_9}d). We consider the motion to be mainly in the 1D vertical dimension where $z(t)$ denotes the displacement of the collophore's hemispherical center with respect to the undisturbed water surface. During impact, hydrodynamic forces are induced by form drag, buoyancy, added mass, surface tension and dissipation through capillary waves \cite{Vella2010,Aristoff2010,Zhaoetal_landing}. We measure the normalized displacement $\bar{z}=(z(t)-z_{t\rightarrow \infty})/D_s$, where $z_{t\rightarrow \infty}$ is the final equilibrium position of the springtail.
 Given the kinematics and physical parameters of the springtails prior to impact, we find that the interplay between capillary and inertial forces determines the dynamics of landing while buoyancy and viscous forces are negligible (See methods and materials and Table S1). We assume that inertial forces are converted into surface deformation and dissipate in the form of capillary waves ($\sim  \gamma D_c \dot z/(g l_c)^{\frac{1}{2}}$). Similar assumptions were used in a previous study on  water striders' landing \cite{Zhaoetal_landing}. To assess the effect of the hydrophilic collophore on the dynamics of springtail landing, we include a capillary adhesion force $F_c = \pi D_c \gamma H(\dot z)$, where $H(\dot z)$ is the heaviside function that acts to retard the movement of the springtail only when it is moving upwards (i.e. when $\dot z$ is positive). The simulations (parameters in Table S2) show that without that additional force, springtail would indeed bounce upwards which qualitatively matches the experimental data. However, the model overestimates the bouncing velocity and trajectory of the springtail which may be attributed to their geometry during impact, as well as to other kinematic parameters such as movement in the x-direction and the rotation of the body. Alternatively, adding the capillary adhesion force $F_c$ rapidly halts the movement of the springtails forcing them to reach its equilibrium position in around $\sim 6\;ms$. It is worth noting that the maximum capillary force calculated by the model is $28\;\mu N$, which is very close to the maximal values calculated in the rotating disk experiments ($\sim 20\;\mu N$). Notice that for the latter, an average mass of 0.13 mg was used, but larger individuals may double that mass. We plot the time series of the forces during impact (Figure \ref{figure_9}c). As expected, surface tension force $F_{ST}$ is dominant, which shows an increase up to a maximum $38$ $\mu N$ as the springtail deforms the water surface. In parallel, the capillary drag $F_{d,c}$ decreases from a maximum of $36$ $\mu N$ to zero when the system changes direction and starts vibrating back. Note that the maximum value of the surface tension force per unit length (wetted collophore perimeter) is $\sim 100\; mN/m$ less than the theoretical force per unit length required to break the surface of water $\sim 144\; mN/m$ \cite{Koh_jumpingonwater}.

\section*{Discussion}

Springtails can perform explosive jumps at the slightest provocation, followed by fast body rotation in mid-air, at rates similar to the wing flapping frequency of insects. All of this in the blink of an eye. Due to the extremity of this escaping behavior, previous researchers have portrayed springtails as unable to control their jumping directionality and aerial rotation, as well as their landing orientation and impact against a surface \cite{christian1978jump, hu2010water, zettel2013jumping}. Here, using experiments and theoretical approaches, we show that in contrary to previous belief, semiaquatic springtails \textit{I. retardatus} can indeed achieve higher levels of locomotion control and maneuvering at all jumping stages, by tuning their impulsive stroke and by modifying their body posture, before and in mid-air. Aerial righting is induced by this U-shaped posture via aerodynamic torque. But more importantly, springtails take advantage of the physical properties of their ventral tube, which confers stable equilibrium and firm adhesion on the water surface, fundamental features for a directed takeoff and favorable landing.

\subsection*{Takeoff control}
Regarding takeoff, our results indicate that springtails can control the launching angle by tilting their body with respect to the water surface. Accordingly, vertical or horizontal jumps occur if the tip of the abdomen is maintained closer or farther from the water surface, respectively. Consequently, the furcula can hit the surface directly or at an angle, producing a more vertical or horizontal impulse. Meanwhile, jumping speed was inversely related with the duration of the furcula stroke, which indicates that the momentum imparted on the water is a function of the stroke rate. On average, horizontal jumps (HDM) were 20$\%$ slower in speed and 30$\%$ lower rotation than those of vertical jumps (VAM). 

We observed that in some cases, springtails skip on the water surface with their collophore remaining anchored. Despite this apparent low performance, individuals moving horizontally were able to travel at speeds of 28 cm/s ($\sim$280 bodies/s) (Figure \ref{figure_3}), which are similar to the speeds of newborn water striders (body size $\sim$1.3 mm) rowing on the water surface \cite{ortega2017escape}. This is intriguing, because sprigtails’ motion depends only on the impulse produced by the forked furcula against the water surface, while for water striders it depends on the reaction force produced by their two elongated middle legs. In the former, during explosive strokes, the momentum transfer seems to be mediated primarily by capillary waves while in the latter by the production of vortical wakes \cite{hu2003hydrodynamics}. In contrast, during horizontal jumps, springtails seem to rely, as water striders do, on the production of vortices shed in the water (Figure \ref{figure_3}). Future investigation on the hydrodynamics of springtails’ jumping is required in order to understand the full details of their momentum transfer. Our results on takeoff control also agree with field studies of springtails during their massive migration over the snow. For example, \textit{Hypogastrura socialis} springtails have been reported to perform oriented jumps and travel up to 300 m/day \cite{hagvar1995long}, which is astonishingly similar to the distance per day covered by the spotted-wing drosophila during dispersal (see \cite{vacas2019survey}), or to the total distance covered in bodies by some migrating ungulates during a whole year \cite{joly2019longest}. How springtails travel at similar rates to those of some animal fliers deserves further investigation. Alternatively, we can use mathematical and robophysical models to understand springtails' jumping behavior. We discover that through changing postures and utilizing their mechanical designs, springtails can obtain control during their takeoff, mid-air adjustment, and landing. This control is not trivial considering their flight speeds, which are on the order of 1000 body lengths per second. It is curious whether other small arthropods can also leverage their physical interactions to control their locomotion.

\subsection*{Collophore's role}

The ventral tube is a unique adaptation for  collembolans. It is used for adhesion, cleaning, drinking and nutrient absorption \cite{chen2019structure}. Furthermore, clover springtails use it as a lever for standing up from an inverted position given that they have a complex and elongated collophore \cite{brackenbury1990novel}. It has been suggested that the ventral tube may be employed in controlling the direction and trajectory during leaping \cite{favret2015adhesive}. In agreement, we found that the collophore enables both a controllable takeoff and landing. During takeoff the collophore, which is anchored to the surface, serves as a stand, making easy pitching or yawing of the body, facilitating a directed and controlled launching. Also, because of this firm adhesion, the collophore enhances the downward transference of the momentum imparted by the furcula. And finally, it can impede its physical detachment from the water if the impulse force is redirected horizontally. In terms of stability, the water collected by the collophore effectively lowers the center of gravity which confers stability, and thus facilitates a ventral landing, as well as a correct standing orientation. Previous experiments indicate that adding extra weights in cockroaches’ legs increases the chances of self-righting \cite{othayoth2021propelling}. This agrees with the results of our experiments of physical models in free fall, as well as with our jumping robot. U-shaped strips with a droplet in the vertex seem to right themselves quicker and with lower angular variation than when they have no droplet. During landing impact, the collophore reattaching to the surface impedes bouncing. And finally, the collophore also produces drag on the water surface, reducing springtails' horizontal speed. We observed that individuals during impact produce capillary waves that seem effective to dissipate their vertical momentum. In contrast, those landing on their backs over the water surface, or even those impacting on its dry collophore on a solid surface, present a dramatic rebound. This bouncing produces a longer posture recover from an unfavorable landing than those approaching ventrally. We estimated that the adhesion force produced by the collophore is $\sim$7$\pm4$  $\mu N$ (max. $\sim$20 $\mu N$) which suggests that water’s capillary forces are used for attachment in springtails.

\subsection*{Aerial righting}
We discovered that springtails, which have no wings, can perform a rapid and effective mid-air maneuver to reduce rotation and gain a favorable orientation. This is the first evidence that non-insect hexapods can perform aerial righting. Surprisingly, springtails can recover from an upside-down posture in less than ~20 ms, which is the fastest aerial righting of any studied wingless animal as far as we know. Free fall geckos, for example, can right themselves in $\sim$100 ms \cite{jusufi2008active}. However, these vertebrates use inertial forces generated by their long tails to rotate in mid-air. Meanwhile, wingless insects such as juvenile stick insects \cite{zeng2017biomechanics} and aphids \cite{ribak2013adaptive}, correct their position from an upside-down fall in ~300 ms and ~200 ms, respectively. In these cases, insects use their elongated legs to produce an aerodynamic torque and thus self-righting \cite{zeng2017biomechanics, ribak2013adaptive}. Plant seeds, such as samaras \cite{ortega2019superb} and dandelion seeds \cite{cummins2018separated}, or even badminton shuttlecocks \cite{hasegawa2013flow}, can passively right themselves by producing an aerodynamic torque, similar to insects. In comparison, we found that springtails adopt a U-shape posture to induce an aerodynamic drag, and at the same time they lower their center of gravity using the water collected by the collophore during takeoff, which enhances a stable equilibrium. Moreover, collembolans seem to use their legs, keeping them spread in mid-air to induce yaw stability. Springtails have elongated antennae that can potentially play some aerodynamic role during righting, however since many individuals oriented them behind their backs after jumping, their effect on righting seems small. On the contrary, other individuals orient the antennae to prolong the body curvature, which can contribute to induce an increased aerodynamic torque. One relevant finding is that springtails can reduce and stop effectively body rotations in mid-air by deforming their bodies. Badminton shuttlecocks, that right themselves using aerodynamic forces, do so by rapidly damping spinning \cite{hasegawa2013flow}, which may be a similar mechanism to that used by springtails. Moreover, we observed that individuals placed in the vertical wind tunnel can perform other complex maneuvers in mid-air. For example, they can brace their legs tight against their bodies and form a ball shape, resembling pillbugs. This produces a rapid vertical descending (Video S3), which may be effective to force landing after being drafted by strong winds. Such extraordinary maneuvering skills exhibited by springtails, ants \cite{yanoviak2010aerial}, bristletails \cite{yanoviak2009gliding} and spiders \cite{yanoviak2015arachnid} require more research attention. Especially, because it can help to understand the origin of flight in insects. Also, notice that the present results may not directly apply to other groups of terrestrial Collembolans, which deserve further research.

\subsection*{Springtail-inspired aerial righting robot}
Several jumping robot designs have been developed based on small wingless jumping animals. For example, a flea-inspired robot (2 cm length, 1 g mass) that uses a catapult mechanism  similar to Siphonapterans can jump vertically ~30 times its body size at speeds of 5 m/s, however this robot does not have the capability to control its orientation during landing \cite{Noh_fleaRobot}. In contrast, our springtail inspired jumping robot, which is ten times lighter, can achieve similar performance to the aforementioned one (Figure \ref{figure8}), but it can right itself using a simple aerodynamic torque, similar to that observed in springtails. A recent robot design based on springtails and midges has the ability to right itself in mid-air \cite{Ma_jumpingRobot}. However this robot is an order of magnitude larger and four orders heavier than ours. Furthermore, the landing control of the later robot is based on lowering the center of gravity relative to its centroid, which confers self-righting by inertia. In contrast, we find that an aerodynamic torque (drag flaps and extra body mass) is an effective mechanism to reduce rotation and gain a ventral position for our small and lighter robots. Future robot designs will be directed to control the timing of deployment of the drag enhancers, as springtails do, to reduce the penalties in maximal height as well as jump from and land on the water surface for the subsequent jumps.                   

\subsection*{Landing on water}

We found that springtails land 85\% of the time on their ventral side. This landing effectiveness is not different from other animals. For example, a recent study in geckos reported that they can right themselves and land successfully on a tree trunk 87\% of the time \cite{siddall2021tails}. Meanwhile, free fall aphids can land on their feet from 60\% to 95\% of the time depending on the dropping height \cite{ribak2013adaptive}. Kinetic energy during the impact is the major concern in animals, because it can cause serious physical damage, or depending on the time of posture recovery, can increase predation risk. Geckos impacting head-first on a vertical trunk roll their bodies counterclockwise and use their long tail and back feet to anchor, which safely and effectively absorbs their kinetic energy \cite{siddall2021tails}. In general, insects use their legs to dampen the momentum during impact. In contrast, springtails are unique in this respect because they use their ventral tube to absorb the impact forces on the water or other solid substrates. Moreover, because the collophore attaches firmly to the surface, it impedes bouncing in the air. Springtails landing on their backs or sides take ten times more time to stop bouncing and stand on their feet than those landing on their collophore (Figure \ref{figure_9}. Video 2). This rapid and effective ventral landing is essential for springtails that live on the air-water interface given the high predation pressure they face especially by underwater, semiaquatic, and terrestrial predators.\\

\subsection*{Concluding remarks}
In conclusion, we found that semiaquatic springtails exert an effective locomotive control at every stage of their jumping. During takeoff they adjust their launching direction and speed by tilting their body and by tuning the duration of the impulsive stroke, respectively, while anchoring to the water using the ventral tube. In mid-air, they decrease body rotation and gain a favorable body orientation during landing by simply curving their bodies, which induces an aerodynamic torque, as well as by lowering their center of gravity using water collected by the collophore. And finally, during impact they produce a capillary wave and anchor to the surface via the collophore, which impedes bouncing. By combining biological experiments, mathematical modeling, and robotic analogs, our results portray springtails, as they should have been in the first place, as highly maneuverable and skillful jumpers, with extraordinary aerodynamic control capabilities. Our work highlights why they are the most abundant and diverse group of living hexapods, just after insects.


\matmethods{
	
Individuals of \textit{I. retardatus} (Figure \ref{figure_1}) were collected at the edge of a large pond close to the Fifth Third Bank Stadium-Kennesaw State University, Kennesaw, GA. We rubbed a plastic cup against dead leaves laying over the water to capture springtails. Then, collected individuals were placed on a large container with some wet leaves to avoid desiccation. Springtails were finally transported to the Bhamla Lab at Georgia Tech, for experimentation. Regarding springtail's identification, preserved specimens were mounted on object slides and studied using an optical microscope (Nikon Eclipse Ti2-U Inverted Research Microscope). Mounting medium used was distilled water.Photographs of diagnostic features were made using a microscope camera (Hamamatsu Flash 4 V3 Camera) and sent to Frans Janssens (Department of Biology, University of Antwerp, Antwerp, Belgium) for identification. Systematics and nomenclature of the taxon can be found online (Bellinger PF, Christiansen KA, Janssens F 1996-2022. Checklist of the Collembola of the World. http://www.collembola.org). Springtails' specimens were deposited at the Georgia Museum of Natural History, University of Georgia. 

\subsection*{Takeoff directional control: kinematics}
In order to understand the leaping kinematics of springtails during the impulsive stroke and takeoff we filmed 27 individuals sideways at 10,000 frames/s using either Cannon MP-E65mm f/2.8 1-5x Macro or 10x microscopic lenses mounted on a Canon EF 200mm f/2.8L II USM. We digitized the tip of the head and the abdomen, the body centroid, and two points along the water level. We calculated instantaneous speed $u_i$ and body angular speed $\omega _i$ using the body centroid and body angle, respectively. For calculating the latter, we used a MSE quintic spline function \cite{walker1998estimating}. 

Using these digitized points, we also calculated body size $l_b$, takeoff angle $\theta$, and height of the tip of the abdomen with respect to the water surface $H_{abdomen}$. Stroke duration of the furcula $t_{furcul}$ was calculated from each sequence as the total frames, from the initial to end of the stroke multiplied by the frame rate. Average stroke angular speed of the furcula $\omega _{furcula}$ was calculated by dividing the change in angle by the furcula stroke duration. Reynolds number ($Re = l_{thick} u_t / v $) during jumping was calculated using the springtail's body thickness ($l_{thick} \sim 0.3 $ mm), jumping speed ($u_t$ $\sim$0.7 m/s) and the kinematic viscosity of air ($v$ $\sim 1.5  \times 10^{-5}$ m$^2$/s). Definitions of the variables extracted can be found in Figure \ref{figure_3}. Dimensions and velocities are normalized by the body length. Angles are in radians. 

\subsection*{Takeoff model} 
We constructed a dynamic model for springtail takeoff. The model obeyed planar rigid-body mechanics, i.e. Newton's 2nd law in the translational and rotational direction. Springtails were abstracted into three rigid parts: its body, collophore, and furcula (Figure \ref{figure_3}c). The proportions of the body parts were obtained from images of the insect. The mass was assumed to be distributed evenly throughout the body portion (0.1 mg), and we used the moment of inertia 0.011 $g \cdot mm^2$, which was obtained through a 3D scan of the springtail. We prescribed the deformation of the body: the furcula opened at a constant angular velocity. The trajectory is determined by the physical forces imposed on the insect, including furcula propulsion, collophore adhesion and gravity.

Furcula force was modelled as the pressure force of a plate moving in the fluid. Due to the high speed of the furcula movement ($\sim$ 1 m/s), fluid inertia dominated over viscous effects. We thus assumed the pressure to be $\frac{1}{2}\rho U_{furcula}^2$, where $\rho$ is water density and $U_{furcula}$ is the normal velocity of furcula moving in the water. Collophore adhesion was modeled based on lubrication theory. The force is equal to $k_c U_{collophore}$, where $U_{collophore}$ is the speed of collophore leaving the water, and $k_c$ is the prefactor that governs the strength of the adhesion. $k_c$ is affected by the thickness of the water film, which we could not accurately measure. We left $k_c$ as a free parameter which we vary in Figure \ref{figure_3}e-g.

We integrated the model using explicit finite different scheme. The simulations were implemented in MATLAB with time steps 0.01 ms long. We varied three parameters: body orientation $\theta$, furcula opening velocity $\omega_{furcula}$, and the collophore coefficient $k_c$. In Figure \ref{figure_3}e-g, to reveal the effect of $\omega_{furcula}$ and $\theta$, we fixed the other variable ($\theta$ and $\omega_{furcula}$, respectively) to the median value of the studied range. Linear regression analysis was performed between kinematic variables.

\subsection*{Collophore’s wetting and adhesion}
Water collected by the collophore during takeoff was estimated from five filmed individuals. We assumed that the collophore was a cylinder with a diameter of 0.12$\pm$0.03 mm and a length of 0.22$\pm$0.06 mm. From video sequences we measured the diameter of the water droplet at the tip of the collophore, which was  0.13$\pm$0.03 mm. Total mass of water was estimated using the water density multiplied by the sum of the volume of the cylinder (collophore) and the volume of the droplet.     

Adhesion force produced by the collophore on a smooth surface was calculated from their maximal transversal acceleration just before alive springtails (n=23) were ejected from the surface of a rotating petri dish (100 mm) (Figure S3). A 1.5-3V 24,000 RPM DC electrical motor connected to a power supply was used to gradually increase the disk’s rotational speed. Individuals were filmed at 1000 frames/s from above. Trajectories of body centroids were digitized and used to calculate speed and acceleration, as it was previously mentioned. Springtail's body mass was estimated as 0.13 mg from weighing 18 individuals at once using an analytical balance. Thus, average adhesion force was calculated from the product of the body mass and the average of the transversal acceleration just before springtails detached from the disc.

\subsection*{Trajectories and landing success}

Kinematics of jumping trajectories and body posture during landing were analyzed as follows. We filmed 20 random individuals, from taking-off to landing, using a FASTCAM SA4 (Photron, Inc) at 10,000 frames/s. Springtails were placed in a plastic container (10×4×10 cm) partially filled with distilled water. The tip of the head and the abdomen, as well as the body centroid were marked using the DLTdv8 program for Matlab \cite{hedrick2008software}. The resulted XY coordinates of these digitized points were used to calculate the body angle and centroid position over time. Instantaneous speed of the centroid ($u_i$) was calculated as described before. We quantified the number of individuals impacting the water surface on their ventral side at angles < 90 degrees. Takeoff angle, rotational frequency, maximal height, and horizontal distance traveled per individual were calculated.

\subsection*{Self-righting in vertical wind tunnel}
A vertical wind tunnel was used to investigate the aerial righting abilities and maneuvering of springtails (Figure \ref{figure_6}a). Flow in the tunnel was generated by two computer fans connected to a speed controller. The tunnel had a contraction ratio of $\sim$2x. A honeycomb was placed on the contracted side of the tunnel to smooth the airflow. A clear plastic tube with a fine mesh on the bottom was fastened to a plastic funnel and placed on the top of the tunnel. Alive and dead springtails were introduced into the clear tube while the fans were off. Subsequently, we turned the fans on and maintained flow speed at around 1 m/s. We filmed sideways the aerial responses of the alive and dead springtails in the vertical flow at 10,000 frames/s. The tip of the head and abdomen, as well as the abdomen's middle point on the ventral side were marked. The duration of posture recovery during self-righting from an upside-down posture was measured for 10 individuals. PIV was used to resolve the velocity field of the flow produced by the wind tunnel. Lycopodium particles were used to seed the flow. A pointer laser (532 nm, 5 mW) was used to produce a two-dimensional vertical laser screen ($\sim$2 mm thickness). Velocity fields were calculated using PIVlab \cite{thielicke2021particle}.

\subsection*{Free fall of physical models}
In order to understand the role of the change in body posture and the droplet collected by the collophore during self-righting we made a free fall experiment using one U-shaped and one flat plastic plate (10.0×0.5 mm). Additionally, we placed a water droplet ($\sim$1 mm diameter) on the vertex of the U-shaped plate to mimic the water collected by the collophore. Thus, the U-shaped (with and without a droplet on its vertex) and the flat plate were released from a height of $\sim$9 cm using tweezers. The orientation of the U-shaped plate during release was approximately upside-down (i.e., its vertex was pointing up), while the flat plate was oriented parallel to the surface. We filmed at 500 frames/s each trial and marked the tips and vertex of the plate. To avoid air flow disturbances, experiments were carried out inside a plastic container (8×4×11 cm). We calculated the orientation angle over time. Average speed of plates in free fall was $\sim$0.65 m/s. Reynolds number was calculated based on the plate thickness.

\subsection*{Bio-inspired robot}

 We designed a small scale jumping robot with a length and mass of ~2 cm and 86 mg, respectively. A smart composite microstructure (SCM) fabrication process was used \cite{wood2008microrobot}. The SCM allows flexure hinge-based folding mechanism without a complex mechanism and assembly. This fabrication process involves laminating 2D planar sheets of multi-materials, and then folding stacked layers into 3D structure to embody a desired shape. The robot comprises of two parts as shown in Figure \ref{figure8}a. The region 1 is a flexure hinge-based composite structure for the rotation of the legs, and region 2 is a single glass-fiber composite layer (FR-4) for energy storage. For aerial righting, we attached a drag flap (Polyimide film) to each end of the robot. Also we added an extra mass in the ventral side that corresponded to 14\% of the body mass ($\sim$12 mg). Both flaps and the extra mass were used to reduce rotational speed and land on the ventral side of the robot via an aerodynamic torque resembling that produced by springtails while deforming the body in mid-air (Figure \ref{figure8}).
 
A torque reversal catapult (TRC) mechanism is used for generating an explosive launching of the robot \cite{Koh_jumpingonwater, steinhardt2021physical}. The jumping mechanism involves four temporal phases driven by energy storage of the elastic beam based on heating of SMA coil actuator as shown in Figure \ref{figure8}. In Phase I, the body and moving linkage are in contact at an angle ($\alpha$), and the robot is in a locked configuration. During Phase II, the contraction of the SMA actuator by heating induces the deformation of the beam, and elastic energy is stored in the structure through deformation. In Phase III, the actuator is continuously heated until the robot reaches the overcentering configuration (i.e., the moving linkage and SMA coil actuator are in a straight line). From Phase I to Phase 3, the SMA actuator and elastic beam involve the deformation until potential energy is at a maximum. In Phase IV, the stored potential energy is converted into kinetic energy of the moving linkage, and the legs attached to the moving linkage rotate at high speed and push the ground away. Figure \ref{figure8} shows the high-speed image of the four phases.

We tested 5 jumps of the robot without any modifications (robot), with the extra mass (robot+m), with the drag flaps (robot+d) and finally with both additions (robot+m+d). Ventral landing success was quantified from 20 jumping trials from a robot with added mass and drag flaps. A Kruskal-Wallis rank sum test was used to test for differences among treatments regarding angular speed, maximal height and number of turns. Post hoc Tukey tests were used for pair-wise multiple comparisons. All data analyses were performed in R \cite{Rbib}.

\subsection*{Impact on the water surface}
To understand how fast springtails recover after landing onto the water surface, we filmed 10 individuals at 10,000 frames/s landing ventrally while attaching on the water surface using their collophore. As a control, we filmed 3 individuals landing on their backs and bouncing. Additionally, we filmed a pair of springtails landing ventrally on a solid surface (plexiglass), but without enough water for the collophore to attach to the surface. For the latter, springtails were kept on a dry plastic container for ~7 minutes without any water available. Using those sequences, we digitized the body centroid and calculated the trajectory, as well as the speed over time. We calculated how long it took from the impact to when the vertical bouncing over the water or solid surface ceased.

\subsection*{Impact on the water surface: Theoretical model}
To analyze the controlled landing of springtails, we developed a reduced order mathematical model inspired by Aristoff \textit{et al.} \cite{Aristoff2010}, Vella \textit{et al.} \cite{Vella2010} and recently by Zhao \textit{et al.} \cite{Zhaoetal_landing}. We observed, that prior to landing, springtails adjust their body posture into a U-shape to land mainly on their collophore (See Video S2). Collophore geometry was assumed as a cylinder with a hemisphere at its end $R_c$ (diameter $D_c$)(see Figure \ref{figure_9}d). We ignored any geometrical effects that may arise due to the arched posture of the springtails as it remains constant during landing. We also assumed that the motion is primarily in 1D in the vertical direction. This approximation is valid since the angle between the velocity vector of the free falling springtail right before impact, and the horizontal free surface of the water is $84.64^{\circ} \pm 3.78\;(N=9)$.
\\
We applied Newton's second law to estimate the motion of the cylinder over time. Upon impact with the surface of water ($\rho=1000\;kg/m^3$, $\mu = 1\;mPa.s$, $\gamma = 72\;mN/m$), a sphere of diameter $D_c$ generates hydrodynamic forces that include form drag $F_{d,f} \propto -\rho D^2\dot z^2$ within the water phase, surface tension $F_s \propto \gamma (S/lc)$ (where $S$ is the perimeter of the collophore), and buoyancy force $F_b \propto \rho g z(D_c)^2$, added mass (inertia) $F_{i} \propto \rho g(D_c)^3$  where $g$ is the gravitational constant and $z$ is the distance between the center of gravity of the cylinder and the undisturbed surface of water, and maximum depth $z_{max}$ is $0.2 \; \pm 0.07\;mm $ (See Figure \ref{figure_9}). In addition, Zhao \textit{et al.} \cite{Zhaoetal_landing} showed that energy dissipation occurs through capillary waves $F_{d,c} \propto (\gamma D_c/c)\dot z$ where $c \sim (gl_c)^{\frac{1}{2}}$ is the speed of the capillary wave and $l_c$ is the capillary length ($\sim 2.7\;mm$ for water).\\
 We write the following equation of motion:\\
 $m_s \ddot z(t) = -m_sg+ F_b+F_{d,c}+F_{d,f}+F_{ST}+F_{i} - F_{collophore} $ \\ 
 where:
 $F_b=(1/4)\rho g \pi {D_c^2} ({D_c}/{3}-z)$, $F_{d,c}=a_1 ({\gamma D_c}/{c}) \dot z H(-\dot z)$, $F_{d,f}=-({1}/{8}) C_d \rho \pi {D^2} \dot z^2 $, $F_{ST}=a_2 \gamma (\pi {D_c}/{l_c}) z$ and $F_i=({1}/{12}) \pi D_c^3 \rho_f $. A list of all parameters is summarized in Table S1. The function $H(\dot z)$ is the heaviside function since we are assuming that energy is mainly dissipated due to capillary waves as the system is traveling downwards, i.e., when the system is below the undisturbed water-air interface. We note that  $a_1 = -1.25$ and $a_2=6$ are pre-factors that were found to best match the experimental data.\\
We estimated the relative contribution of these hydrodynamic forces on the impact dynamics through dimensional analysis. Assuming that the springtails do not penetrate the surface of water, we find that given their size ($L\sim 1.5 \times 10^{-3}, D_s\sim 0.5 \times 10^{-3}, D_c \sim 0.125 \times 10^{-3}$) and impact velocity ($u\sim 0.5-1\;m/s$), surface tension force is dominant as summarized by table S2. This can also be shown by taking the force ratios which scale as ${F_b}/{F_{ST}} \sim 10^{-3} $, ${F_{d,f}}/{F_{ST}} \sim 10^{-1}$ ${F_i}/{F_{ST}} \sim 10^{-3}$.\\
The equation of motion is solved numerically using the 4th order Runga-Kutta solver (ode45) in MATLAB.\\
To simplify the equations (mainly the buoyancy and form drag equations), the systems starts at $t=0$ when the hemisphere is submerged in water ($z_o=0$) with a downward velocity equal to the impact velocity of $u=0.7\;m/s$. To assess the role of collophore adhesion, we solve these equations of motion with and without the added capillary adhesive force of the collophore $F_c$. In this case,  $F_{c}=\pi D_c \gamma H (\dot z)$ where $D_c$ is the collophore diameter and $H (\dot z)$ is the heaviside function since capillary adhesion is activated only when the body is traveling upwards.

}
\showmatmethods{} 

\acknow{We thank Sunghwan (Sunny) Jung for comments and suggestions; Frans Janssens for springtail species identification; Edward R. Hoebeke for allocating voucher specimens of springtails at the Georgia Museum of Natural History; and  members of the Bhamla lab for feedback and useful discussions. MSB acknowledges the funding support from NIH Grant R35GM142588, NSF Grants CAREER 1941933 and 1817334, and gift funding from the Open Philanthropy Project. JSK acknowledges the support from the Ajou University research fund and Basic Science Research Program through the National Research Foundation of Korea (NRF-2021R1C1C1011872)}. 

\showacknow{} 

\bibliography{refs}

\end{document}